\documentclass{aastex63}

\received{September 9, 2019}
\revised{}
\accepted{October 21, 2019}
\submitjournal{AJ}

\shorttitle{Atmospheric constraint of (50000) Quaoar}
\shortauthors{Arimatsu et al.}

\begin{document}

\title{New constraint on the atmosphere of (50000) Quaoar from a stellar occultation}

\correspondingauthor{Ko Arimatsu}
\email{arimatsu@kwasan.kyoto-u.ac.jp}

\author[0000-0003-1260-9502]{Ko Arimatsu}
\affiliation{Astronomical Observatory, Graduate School of Science,  Kyoto University \\
Kitashirakawa-oiwake-cho, Sakyo-ku, \\
Kyoto 606-8502, Japan}

\author[0000-0001-5797-6010]{Ryou Ohsawa}
\affiliation{Institute of Astronomy, Graduate School of Science, The University of Tokyo, 2-21-1 Osawa, Mitaka, Tokyo 181-0015, Japan}

\author[0000-0002-3821-6881]{George L. Hashimoto}
\affiliation{Department of Earth Science, Okayama University, 1-1-1 Kita-ku Tsushimanaka, Okayama 700-8530, Japan}

\author[0000-0001-7501-8983]{Seitaro Urakawa}
\affiliation{Japan Spaceguard Association, Bisei Spaceguard Center 1716-3 Okura, Bisei, Ibara, Okayama 714-1411, Japan}

\author{Jun Takahashi}
\affiliation{Center for Astronomy, University of Hyogo 407-2 Nishigaichi, Sayo, Hyogo 679-5313, Japan}

\author{Miyako Tozuka}
\affiliation{Center for Astronomy, University of Hyogo 407-2 Nishigaichi, Sayo, Hyogo 679-5313, Japan}

\author{Yoichi Itoh}
\affiliation{Center for Astronomy, University of Hyogo 407-2 Nishigaichi, Sayo, Hyogo 679-5313, Japan}

\author{Misato Yamashita}
\affiliation{Department of Earth Science, Okayama University, 1-1-1 Kita-ku Tsushimanaka, Okayama 700-8530, Japan}

\author[0000-0003-2273-0103]{Fumihiko Usui}
\affiliation{Center for Planetary Science, Graduate School of Science, Kobe University, 7-1-48 Minatojima- Minamimachi, Chuo-Ku, Kobe, Hyogo 650-0047, Japan}

\author{Tsutomu Aoki}
\affiliation{Kiso Observatory, Institute of Astronomy, School of Science, The University of Tokyo 10762-30, Mitake, Kiso-machi, Kiso-gun, Nagano 397-0101, Japan}

\author{Noriaki Arima}
\affiliation{Institute of Astronomy, Graduate School of Science, University of Tokyo, 2-21-1, Osawa, Mitaka, Tokyo 181-0015, Japan}

\author{Mamoru Doi} 
\affiliation{Institute of Astronomy, Graduate School of Science, University of Tokyo, 2-21-1, Osawa, Mitaka, Tokyo 181-0015, Japan}
\affiliation{Research Center for the Early Universe, Graduate School of Science, The University of Tokyo, 7-3-1 Hongo, Bunkyo-ku, Tokyo 113-0033, Japan}

\author{Makoto Ichiki}
\affiliation{Institute of Astronomy, Graduate School of Science, University of Tokyo, 2-21-1, Osawa, Mitaka, Tokyo 181-0015, Japan}

\author{Shiro Ikeda}
\affiliation{The Institute of Statistical Mathematics, 10-3 Midori-cho, Tachikawa, Tokyo 190-8562, Japan}

\author{Yoshifusa Ita}
\affiliation{Tohoku University, 6-3 Aramaki, Aoba, Aoba-ku, Sendai, Miyagi 980-8578, Japan}

\author{Toshihiro Kasuga}
\affiliation{National Astronomical Observatory of Japan, 2-21-1 Osawa Mitaka Tokyo 181-8588 Japan}
\affiliation{Department of Physics, Kyoto Sangyo University, Motoyama Kamigamo Kita-ku Kyoto 603-8555 Japan}

\author{Naoto Kobayashi}
\affiliation{Institute of Astronomy, Graduate School of Science, University of Tokyo, 2-21-1, Osawa, Mitaka, Tokyo 181-0015, Japan}
\affiliation{Kiso Observatory, Institute of Astronomy, School of Science, The University of Tokyo 10762-30, Mitake, Kiso-machi, Kiso-gun, Nagano 397-0101, Japan}

\author{Mitsuru Kokubo}
\affiliation{Tohoku University, 6-3 Aramaki, Aoba, Aoba-ku, Sendai, Miyagi 980-8578, Japan}

\author{Masahiro Konishi}
\affiliation{Institute of Astronomy, Graduate School of Science, University of Tokyo, 2-21-1, Osawa, Mitaka, Tokyo 181-0015, Japan}

\author{Hiroyuki Maehara}
\affiliation{Okayama Branch Office, Subaru Telescope, National Astronomical Observatory of Japan, NINS, Kamogata, Asakuchi, Okayama, Japan}

\author{Noriyuki Matsunaga}
\affiliation{Department of Astronomy, Graduate School of Science, The University of Tokyo, 7-3-1 Hongo, Bunkyo-ku, Tokyo 113-0033, Japan}

\author{Takashi Miyata}
\affiliation{Institute of Astronomy, Graduate School of Science, University of Tokyo, 2-21-1, Osawa, Mitaka, Tokyo 181-0015, Japan}

\author{Mikio Morii}
\affiliation{The Institute of Statistical Mathematics, 10-3 Midori-cho, Tachikawa, Tokyo 190-8562, Japan}

\author{Tomoki Morokuma}
\affiliation{Institute of Astronomy, Graduate School of Science, University of Tokyo, 2-21-1, Osawa, Mitaka, Tokyo 181-0015, Japan}

\author{Kentaro Motohara}
\affiliation{Institute of Astronomy, Graduate School of Science, University of Tokyo, 2-21-1, Osawa, Mitaka, Tokyo 181-0015, Japan}

\author{Yoshikazu Nakada}
\affiliation{Institute of Astronomy, Graduate School of Science, University of Tokyo, 2-21-1, Osawa, Mitaka, Tokyo 181-0015, Japan}

\author{Shin-ichiro Okumura}
\affiliation{Japan Spaceguard Association, Bisei Spaceguard Center 1716-3 Okura, Bisei, Ibara, Okayama 714-1411, Japan}

\author{Shigeyuki Sako}
\affiliation{Institute of Astronomy, Graduate School of Science, University of Tokyo, 2-21-1, Osawa, Mitaka, Tokyo 181-0015, Japan}

\author{Yuki Sarugaku}
\affiliation{Department of Physics, Kyoto Sangyo University, Motoyama Kamigamo Kita-ku Kyoto 603-8555 Japan}

\author{Mikiya Sato}
\affiliation{The Nippon Meteor Society}

\author{Toshikazu Shigeyama}
\affiliation{Research Center for the Early Universe, Graduate School of Science, The University of Tokyo, 7-3-1 Hongo, Bunkyo-ku, Tokyo 113-0033, Japan}

\author{Takao Soyano}
\affiliation{Kiso Observatory, Institute of Astronomy, School of Science, The University of Tokyo 10762-30, Mitake, Kiso-machi, Kiso-gun, Nagano 397-0101, Japan}

\author{Hidenori Takahashi}
\affiliation{Institute of Astronomy, Graduate School of Science, University of Tokyo, 2-21-1, Osawa, Mitaka, Tokyo 181-0015, Japan}
\affiliation{Kiso Observatory, Institute of Astronomy, School of Science, The University of Tokyo 10762-30, Mitake, Kiso-machi, Kiso-gun, Nagano 397-0101, Japan}

\author{Ken'ichi Tarusawa}
\affiliation{Kiso Observatory, Institute of Astronomy, School of Science, The University of Tokyo 10762-30, Mitake, Kiso-machi, Kiso-gun, Nagano 397-0101, Japan}

\author{Nozomu Tominaga}
\affiliation{Department of Physics, Faculty of Science and Engineering, Konan University, 8-9-1 Okamoto, Kobe, Hyogo 658-8501, Japan}
\affiliation{Kavli Institute for the Physics and Mathematics of the Universe (WPI), The University of Tokyo, 5-1-5 Kashiwanoha, Kashiwa, Chiba 277-8583, Japan}

\author{Jun-ichi Watanabe}
\affiliation{National Astronomical Observatory of Japan, 2-21-1 Osawa Mitaka Tokyo 181-8588 Japan}

\author{Takuya Yamashita}
\affiliation{National Astronomical Observatory of Japan, 2-21-1 Osawa Mitaka Tokyo 181-8588 Japan}

\author{Makoto Yoshikawa}
\affiliation{Japan Aerospace eXploration Agency, 3-1-1 Yoshinodai, Chuo-ku, Sagamihara, Kanagawa 252- 5210, Japan}

\begin{abstract}
We report observations of a stellar occultation by the classical Kuiper belt object (50000) Quaoar occurred on 28 June 2019.
A single-chord high-cadence (2~Hz) photometry dataset was obtained with the Tomo-e Gozen CMOS camera mounted on the 1.05 m Schmidt telescope at Kiso Observatory. 
The obtained ingress and egress data do not show any indication of atmospheric refraction and allow to set new $1\sigma$ and  $3\sigma$ upper limits of  6 and 16 nbar, respectively, for the surface pressure of a pure methane atmosphere.
These upper limits are lower than the saturation vapor pressure of methane at Quaoar's expected mean surface temperature 
($T \sim 44$~K)
and 
imply the absence of a $\sim$10 nbar-level global atmosphere formed by methane ice on Quaoar's surface.
\end{abstract}


\section{Introduction} \label{sec:intro}

Large sized trans-Neptunian objects (TNOs) are known to retain volatile ices on their surfaces. 
These ices are thought to sublime and possibly form tenuous atmospheres.
The loss of volatiles to space through the atmosphere appears to be a primary process that accounts for the observed diversity of the TNOs' surface composition \citep{Schaller07b}.
The presence of the atmospheres may also contribute to resurfacing processes possibly acting on several TNOs, 
whose surface ages deduced from spectroscopic observations are much younger than the age of the solar system \citep{Jewitt04}.
Therefore detection and characterization of TNOs' atmospheres should be a key to understand their chemical and geological evolutions.
Stellar occultation observations are a powerful tool to explore  tenuous atmospheres of TNOs.
The vertical refractivity variations of an atmosphere surrounding a distant body produce clear refraction features in the light curve of an occulted star. 
Since the characteristic timescale of the features on the light curves is expected to be $\sim 1$ s (e.g., \citealt{Ortiz12}), 
high cadence stellar occultation observations are thus required to detect and investigate TNO's tenuous atmospheres.

(50000) Quaoar is a classical Kuiper belt object discovered by \citet{Brown04}.
Its orbital semi-major axis (43.6 au), inclination (8\degr.0), and eccentricity (0.038) are typical for classical Kuiper belt objects.
Quaoar's radius is well studied by previous observations, especially by stellar occultation observations;
the equivalent and equatorial radii estimated from multi-chord stellar occultation observations are $R_{\rm equi} = 555 \pm 2.5~{\rm km}$ and $R_{\rm equa} = 569^{+24}_{-17}~{\rm km}$, respectively \citep{Braga-Ribas13}.
Since it has one known satellite, Weywot \citep{Brown07}, Quaoar's mass ($1.3 - 1.5 \times 10^{21}~{\rm kg}$) was measured from the satellite's orbital motion \citep{Fraser13a}.
The relatively large size and mass indicate that Quaoar is possibly a dwarf planet. 
On the near-infrared reflectance spectrum of Quaoar, there are absorption features attributed to crystalline ${\rm H_2 O}$ and ${\rm NH_3 \cdot H_2 O}$ ices \citep{Jewitt04}.
Since both the volatile ices on TNOs' surface should be destroyed by cosmic rays and solar wind particles on a timescale of $\sim 10^7$ years,
their results suggest resurfacing processes currently acting on Quaoar's surface. 
Near-infrared spectral studies also reported the possible presence of ${\rm CH_4}$ \citep{Schaller07a}, ${\rm C_2 H_6}$ \citep{DalleOre09} and ${\rm N_2}$ ices \citep{Barucci15},
and indicated that Quaoar is a volatile-rich body.
Constraints on the presence of Quaoar's atmosphere have been obtained by stellar occultation observations, such as multi-chord occultation observations by \citet{Braga-Ribas13} and investigations of a close passage of a star by \citet{Fraser13b}.
These studies either found no clear evidence for a global atmosphere around Quaoar. 
However, according to the upper limit on the surface pressure based on their results, they could not rule out the possibility of a tenuous ${\rm CH_4}$ atmosphere equilibrated with surface ${\rm CH_4}$ ice. 
Further investigations were required for understanding the atmosphere of Quaoar.

In this paper, we report the observation of a stellar occultation by Quaoar on 28 June 2019.
High-cadence ($2~{\rm Hz}$) photometric data of the stellar occultation by Quaoar have been obtained with the Tomo-e Gozen camera \citep{Sako18} mounted on the 1.05 m Schmidt telescope at Kiso Observatory.
Tomo-e provides a unique opportunity to carry out high time resolution observations of stellar occultations
and allows explorations of extremely tenuous atmospheres on TNOs.
Section 2 presents the outline of the stellar occultation observation with Tomo-e
and the data reduction method to derive the light curve.
The results of the observation are given in Section 3.
Finally, we summarize the conclusion and prospects in Section 4.

\section{Observation And Data Reduction} \label{sec:obs}

\subsection{Observation}

The occultation of a faint star (Gaia DR2 catalog source ID: 4145978492632029696, ICRS position at epoch J2000: $\alpha =  18^{\rm h}~08^{\rm m}~15^{\rm s}.5$, $\delta = -15\degr~17'~58"$, Gaia Gmag = 15.73; \citealt{Gaia18}) on 28 June 2019 was originally predicted by ERC Lucky Star project\footnote{\url{http://lesia.obspm.fr/lucky-star/occ.php?p=16559}}. %,
The project predicts stellar occultations by TNOs using {\it Numerical Integration of the Motion of an Asteroid} (NIMA) version 3 orbital elements \citep{Camargo14,Desmars15}. 
The prediction confirmed that Quaoar's shadow swept over the major islands of Japan at a relative velocity $v_{\rm rel} = 24.6~{\rm km~s^{-1}}$. 
This $v_{\rm rel}$ and Quaoar's radius ($R_{\rm equi} = 555 \pm 2.5~{\rm km}$; \citealt{Braga-Ribas13}) yielded a maximum possible duration of the occultation to be $45.1$~s.
At the geocentric distance of Quaoar, $D = 41.85~{\rm au}$, the angular diameter of Quaoar during the occultation corresponds to $\sim 37$ milliarcseconds.
The same occultation event was also predicted by {\it Research and Education Collaborative Occultation Network} (RECON) project\footnote{\url{https://www.boulder.swri.edu/~buie/recon/events/50000_190628_0316725.html}}, which is a coordinated TNO occultation observation network \citep{Buie16}.

On 28 June 2019, we carried out observations of the occultation event at four stations distributed in Japan, as summarized in Table 1. 
However,  
only one out of the four stations, Kiso Observatory, could acquire data under a good weather condition.
At Kiso, photometric data were obtained using the Tomo-e Gozen camera \citep{Sako18} mounted on a 1.05~m $f$/3.1 Schmidt telescope.
Tomo-e is an optical wide-field camera developed for time-domain astronomy, equipped with 84 CMOS sensors developed by Canon Inc.
The sensitive wavelength range of the CMOS sensor is  350 to 700 nm with a peak at $\sim 500$ nm \citep{Kojima18}.
The field of view (FoV) for each CMOS sensor is 39\arcmin.7 x 22\arcmin.4  with an angular pixel scale of 1.\arcsec19. 
This Tomo-e Gozen camera offers a sequential shooting mode at a maximum frame rate of 2~Hz (= 0.5~s integration) with rolling shutter.
According to the results of performance tests by \citet{Sako18}, the absolute timing uncertainty of the Tomo-e Gozen camera was less than 0.1 milliseconds.

The target star was observed with the Tomo-e Gozen camera in a 15-minute window centered on the predicted central time of the occultation
(12h 37m on 28 June 2019 UTC).
The exposure time was 0.5~s for each frame corresponding to the frame rate of $\sim 2.0~{\rm Hz}$.
The readout overhead time between exposures was less than 0.1 milliseconds.
At the relative velocity of Quaoar, $v_{\rm rel} = 24.6~{\rm km~s^{-1}}$, the time resolution corresponds to a physical resolution of $12.3~{\rm km}$ on Quaoar's surface.
No filter was used in the present observation.
A movie dataset consisting of 180 frames were compiled into a FITS data cube by one capture procedure.
The time interval between the individual data cubes' integrations was approximately 3.6~s.
During the observation, 10 data cubes (corresponding to 1800 frames in total) were obtained by each sensor. 
In the following analyses, we used the data cubes obtained by a single sensor that detected the occulted star.

\subsection{Data reduction}

Aperture photometry for the occulted star (plus the occulting body) is performed using the data cube after bias and flat-field corrections.
We should note that Quaoar itself was not detected from the obtained data during the occultation with a $3\sigma$ upper limit G-band magnitude of 18.8. 
A constant sky background value was estimated from the average value of the aperture fluxes obtained during the occultation and was subtracted from the light curve.
Due to possible rapid changes in atmospheric conditions, 
the obtained flux values for individual data points can suffer time-dependent flux fluctuations.
In order to correct the possible flux fluctuations and calibrate the flux value of the target star, 
we carry out differential photometry with 148 reference stars (Gaia DR2 G-band magnitudes of $12 - 15$) simultaneously detected in the same sensor.
The calibrated light curve is then normalized to the unocculted flux. 
Figure~1 shows the light curve of the target star after the calibration and the normalization.
Except for the Quaoar's occultation,  
no stellar occultation candidate event by Quaoar's satellite Weywot or by other unknown satellites or rings was found in the light curve.

\section{Results}
\subsection{Length of the observed chord}

As shown in Figure~1, the observed duration of the occultation is roughly $\sim 45$~s,
which is comparable to the maximum possible duration of the occultation ($45.1$~s, see Section~2.1). 
In order to measure the precise physical length of the observed chord,
we fit to ingress and egress part of the light curve 
with a sharp edge occultation model convolved by Fresnel diffraction, 
the angular diameter and the spectral energy distribution of the occulted star,
and the efficiency and the finite integration time of the Tomo-e CMOS sensor.

A characteristic scale of the Fresnel diffraction effect is defined as the Fresnel scale, $L_{\rm F} = \sqrt{\lambda D / 2}$,
where $\lambda$ is the wavelength of observation and $D$ is the distance between the observer and Quaoar.
$L_{\rm F}$ of the observed occultation event thus varies from 1.0 to $1.5~{\rm km}$ at the distance of Quaoar, $D = 41.85~{\rm au}$,
when $\lambda$ varies from 350 to 700 nm.
To estimate the effects of diffraction, 
we generate Fresnel diffraction profiles for a circular occulting disk using a numerical code originally developed for the simulation of the stellar occultation light curve by TNOs \citep{Arimatsu17, Arimatsu19}.
In this calculation, we assume that Kiso Observatory's chord sampled almost the central part of Quaoar 
and the observed occultation direction can be approximated to be perpendicular to Quaoar's limb as our baseline case. 
Since the Tomo-e Gozen camera was used in no filter mode, both the spectral efficiency of the camera and the stellar spectrum should be taken into account for producing synthetic light curves.
Furthermore, the profiles should be convolved by the finite size of the occulted star.  
To estimate the spectrum and the angular diameter of the occulted star,
we carry out a spectral model fit with the following flux catalogue data:
Gaia DR2 \citep{Gaia18} BP ($m_{\rm BP} = 17.35 \pm 0.01$) and RP ($m_{\rm RP} = 14.468 \pm 0.004$) bands and 2MASS \citep{Skrutskie06} J ($m_{\rm J} = 12.18 \pm 0.04$), H ($m_{\rm H} = 11.09 \pm 0.04$) and Ks ($m_{\rm K} = 10.73 \pm 0.03$) bands. 
The comparison of these cataloged values with a stellar spectrum model \citep{Castelli03} indicates 
that the cataloged fluxes are explained by spectral models with an effective temperature and $\log[g]$ of $T_{\rm eff} = 3750~{\rm K}$ and $\log[g] = 2.0$, respectively. 
We thus select a spectral model with $T_{\rm eff} = 3750~{\rm K}$, $\log[g] = 2.0$, and $[M/H] = 0$ (solar metallicity) as the template stellar spectral energy distribution. 
$T_{\rm eff}$ for the selected model is consistent with that from the Gaia DR2 catalog ($T_{\rm eff} = 4027^{+582}_{-742}~{\rm K}$; \citealt{Gaia18}).
The best-fit parameters yield a stellar angular diameter of  $0.0417 \pm 0.0007$ milliarcseconds and an extinction value of $E(\bv) = 1.88 \pm 0.02$. 
At the geocentric distance of Quaoar ($D = 41.85~{\rm au}$), the best-fit angular diameter corresponds to 1.3~km in the celestial plane.
With the best-fit parameters, selected stellar spectral model and system efficiency \citep{Kojima18} and the integration time of the Tomo-e Gozen camera,
the diffraction profiles are convolved to generate a synthetic light curve directly comparable to the observed data points.

The synthetic light curve is fitted to the data points obtained in 6.6~s windows centered on the apparent ingress/egress time of the occultation by minimizing $\chi^2$; \begin{equation}
\label{eq_chi}
\chi^2 = \sum_i \frac{(\phi_{i, {\rm obs}} - \phi_{i, {\rm syn}})^2}{\sigma_i^2},
\end{equation}
where $\phi_{i, {\rm obs}}$, $\phi_{i, {\rm syn}}$, and $\sigma_i$ are the observed and synthetic stellar fluxes, and the $1\sigma$ error at a data point $i$, respectively.
The best-fit results are shown in Figure~2. 
From the light curve fitting, we obtained an event duration of $45.50 \pm 0.05$ seconds. 
Considering a velocity of $v_{\rm rel} = 24.6~{\rm km~s^{-1}}$, it is equivalent to a chord length of $1121.0 \pm 1.2 ~{\rm km}$.
The best-fit $\chi^2$ corresponds to 37.0 with 25 degrees of freedom.
The center-to-edge distance of the occultation (corresponding to one half of the best-fit chord length; $560.5 \pm 0.6 ~{\rm km}$) is slightly larger than the equivalent radius $R_{\rm equi} = 555 \pm 2.5~{\rm km}$ and comparable to the equatorial radius $R_{\rm equa} = 569^{+24}_{-17}~{\rm km} \ $ estimated by the previous occultation study \citep{Braga-Ribas13}.
This result is thus consistent with our baseline assumption that the observed chord sampled the central part. 
However, this single chord does not bring further constraints on the Quaoar's shape.
The central time of the occultation derived from the fit is 12:37:17.79 $\pm 0.02$ on 28 June 2019 UTC.
We should note that the effect of the diffraction and the selection of the stellar parameters do not significantly 
affect the fit results
because the resultant synthetic light curve is largely dominated by the finite integration time.

\subsection{Constraints on atmosphere}

In order to make constrains on Quaoar's atmosphere from the light curve data, 
synthetic light curves are calculated with an atmospheric refraction model based on a ray trace technique described by \citet{Sicardy99}.
The atmospheric refraction model requires an assumption on its composition.
Although the previous occultation studies by \citet{Braga-Ribas13} and \citet{Fraser13b} ruled out the existence of Quaoar's isothermal ${\rm N_2}$ or CO atmosphere due to the lack of corresponding refraction features in the observed light curves, they cannot rule out the possibility of a tenuous ${\rm CH_4}$ atmosphere since its refraction feature in sublimation equilibrium is expected to be much fainter.
The possibility of a ${\rm CH_4}$ atmosphere is also consistent with the existence of  ${\rm CH_4}$ ice feature on the near-infrared spectrum of Quaoar's surface \citep{Schaller07a}.
We thus consider a pure ${\rm CH_4}$ atmosphere. 
For a vertical temperature profile of Quaoar's pure ${\rm CH_4}$ atmosphere, we assume a Pluto-like profile, based on the previous study by \citet{Braga-Ribas13}.
Heating of ${\rm CH_4}$ molecules by their absorption of solar near-infrared radiation should increase the temperature at the stratosphere.
We thus adopt a temperature profile increasing from an equilibrium surface temperature (altitude $z = 0 \, {\rm km}$)
to $z \sim ~10$~km with a gradient of $\mathrm{d}T/\mathrm{d}z =5.7~{\rm K\, km^{-1}}$ and isothermal stratosphere of  $T = 102~{\rm K}$, as proposed in \citet{Braga-Ribas13}.
For the surface temperature, we adopt $T = 44~{\rm K}$, which is the mean surface temperature measured from mid- to far-infrared observations \citep{Fornasier13} and is comparable to the upper limit temperature obtained by the previous occultation study (44.3~K, \citealt{Fraser13b}).
On the other hand, \citet{Fraser13b} pointed out that a slight uncertainty of the assumed surface temperature could cause a large variation in the observational constraint on the surface pressure.
Thus we also adopt  $T = 42~{\rm K}$, corresponding to the equilibrium surface temperature assumed by \citet{Braga-Ribas13}, as a reference. 

Since we assume a pure molecular ${\rm CH_4}$ atmosphere, the atmospheric refractivity $\nu(z)$ at an altitude $z$ is given by
\begin{equation}
\nu(z) = n(z) \times K_{\rm CH_4},
\end{equation}
where $n$ and $K_{\rm CH_4}$ are the number density at $z$ and the molecular refractivity of ${\rm CH_4}$, respectively. 
We take $K_{\rm CH_4} = 1.629 \times 10^{-23}~{\rm cm^{3}~molecular^{-1}}$ given by \citet{Bose72}.
The number density $n(z)$ is calculated by integrating the hydrostatic equation from the surface upwards, assuming that the surface pressure is $p_s$ and it is an ideal gas,  
\begin{equation}
\frac{\mathrm{d}p(z)}{\mathrm{d}z} = - \mu\,n(z)\,g,
\label{hydro}
\end{equation}
where $\mu$ is the mean molecular mass and $g$ is the gravitational acceleration.
In equation~\ref{hydro}, we assume $g = 0.4~{\rm m ~s^{-2}}$ given by \citet{Braga-Ribas13}.
$p(z)$ is the pressure at $z$ and $p(0) = p_s$.
For a stellar ray with an impact parameter from Quaoar's shadow center $r$, its bending angle $\omega(r)$ is calculated by
\begin{equation}
\omega(r) = \int^{+\infty}_{-\infty}  \frac{r}{x}  \frac{\partial \nu}{\partial r}~\mathrm{d}x,
\end{equation}
where $x$ is the trajectory of the ray.
For simplicity, we assume a spherical shape of Quaoar in the model.
The apparent distance of the ray from the shadow center is thus $r + \omega(r) \times D$, where $D$ is the distance of Quaoar ($D = 41.85~{\rm au}$).
Since $\omega(r)$ tends to be less than zero, 
the atmospheric refraction could cause reducing the apparent radius of the occulting object.
We thus generate synthetic light curves for different radii and calculated their $\chi^2$ values derived by equation~\ref{eq_chi}. 
In order to derive the synthetic light curve, each atmospheric refraction profile is convolved by the stellar angular diameter and the finite integration time.
Figure~3 shows the distribution of the obtained $\chi^2$ values for different radius models against the surface pressure $p_{\rm s}$. 
The $\chi^2$ values for individual radii show local minimums at different radii.
However, the envelope of these $\chi^2$ curves reaches its minimum value at $p_{\rm s} = 0$~nbar.
According to the $\chi^2$ envelope for the $T = 44~{\rm K}$ models, 
we find  $1\sigma$ and $3\sigma$ detection upper limits are  $p_{\rm s} = 6$ and $p_{\rm s} = 16$~nbar, respectively.
On the other hand, the $T = 42~{\rm K}$ models give $1\sigma$ and $3\sigma$ upper limits of $p_{\rm s} = 5$ and $p_{\rm s} = 15$~nbar, respectively.
In the present analysis, the obtained upper limit does not significantly depend on the uncertainty of the assumed surface temperature 
since the synthetic light curve is predominantly determined by $p_{\rm s}$ in the assumed ${\rm CH_4}$ atmosphere model.
Figure 4 shows the observed ingress and egress profiles overlaid with the synthetic light curves of the $T = 44~{\rm K}$ models with $1\sigma$ and $3\sigma$ upper limit $p_{\rm s}$ values.

As already described in Section~1, the existence of Quaoar's atmosphere has been discussed in the previous occultation studies.
The upper limit on $p_{\rm s}$ of Quaoar is significantly lower than the previous measurements by \citet{Braga-Ribas13} ($1\sigma$ and $3\sigma$ upper limits of  $p_{\rm s} = 21~{\rm and}~56~{\rm nbar}$, respectively, assuming $T = 42~{\rm K}$) and \citet{Fraser13b} (a $3\sigma$ upper limit surface pressure and temperature of  $p_{\rm s} = 138~{\rm nbar}$ and $T = 44.3~{\rm K}$, respectively).
Since the saturation vapor pressure of ${\rm CH_4}$ at $T = 44~{\rm K}$ and $T = 42~{\rm K}$ are $120$ and $33~{\rm nbar}$, respectively \citep{Fray09}, 
the obtained upper limit pressure is smaller than the expected surface pressure of a possible global atmosphere equilibrated with ${\rm CH_4}$ ice.
However, the saturation vapor pressure is highly dependent on the surface temperature. 
A $p_{\rm s} = 6$~nbar ${\rm CH_4}$ atmosphere would imply an equilibrium temperature of 39.6~K \citep{Fray09}, 
which is lower than but not significantly different from the mean surface temperature ($T = 44$~K, \citealt{Fornasier13}).
If the surface temperature is lower than 39.6~K, 
the presence of an extremely tenuous atmosphere in equilibrium with ${\rm CH_4}$ ice 
is consistent with our non-detection at the obtained upper limits.
Finally, our single-chord data cannot rule out the possibility of a local atmosphere suggested by \citet{Braga-Ribas13}.
According to the previous study of local atmospheres on TNOs \citep{Ortiz12}, 
a focusing effect of a local atmosphere with a surface pressure greater than $\sim {\rm microbars}$ could cause 
"a central flash"; a short-timescale brightening of an occulted star that occurs in the middle of an occultation.
We should note that, however, any signature of the central flash candidate is found in our 2~Hz cadence light curve during the occultation with a $2\sigma$ upper limit of $11\%$ of the total stellar flux. 
Since thermophysical properties of Quaoar are uncertain,
it is difficult to put a limit on a local atmosphere in the present study.
Further multi-chord occultation observations, as well as investigations of the surface thermal properties, will provide valuable information on the atmospheric conditions of Quaoar.

\section{Conclusion and future prospects}
Based on the high-cadence stellar occultation observation made with Kiso Observatory's Tomo-e Gozen camera, 
we have obtained the upper limits on the surface atmospheric pressure of (50000) Quaoar,
which are significantly lower than the previous measurements.
The measured upper limits are smaller than the expected vapor pressure of methane at Quaoar's mean equilibrium surface temperature 
and imply the absence of a $\sim $ 10 nbar-level global atmosphere formed by methane ice on Quaoar's surface.

The present results demonstrate the potential of high-cadence observations 
of stellar occultations achieved by high-speed CMOS cameras.
Our high-cadence data obtained by the Tomo-e Gozen camera provide a unique opportunity to establish new upper limits on the surface pressure of Quaoar
since the timescale of the refraction features produced by extremely tenuous atmospheres on the light curves is expected to be only $\sim 1~{\rm s}$.
Future stellar occultation observations with Tomo-e and other high-speed CMOS cameras will lead to further information about atmospheric conditions around TNOs.

\clearpage

\begin{deluxetable*}{ccccc}[b!]
\tablecaption{ Circumstances of the 28 June 2019 stellar occultation observations for observing stations \label{tab:obs}}
\tablecolumns{5}
\tablenum{1}
\tablewidth{0pt}
\tablehead{
\colhead{Observation site} & \colhead{Longitude}  & \colhead{Aperture}               & \colhead{Obs. Status} & \colhead{Observers}\\
\colhead{}                            &  \colhead{Latitude}            & \colhead{Camera}               & \colhead{}                    & \colhead{}   \\      
\colhead{}                            &  \colhead{Altitude}            & \colhead{}               & \colhead{}                    & \colhead{}   \\  
}
\startdata 
Kiso  Observatory               & $137\degr$ $37'$ $31.5"$ E     &   $1.05~{\rm m}$          &     Positive                   &    R. Ohsawa    \\
                                            & $35\degr$ $47'$ $50.0"$ N   &   Tomo-e Gozen         &                                   &                            \\
                                            &  1132~m                                   &                                    &                                   &                           \\
                                              &                                               &                                   &                                   &                             \\
 Nishi-Harima Astronomical Observatory   & $134\degr$ $20'$ $08.0"$ E  &   $2.0~{\rm m}$           &       Bad weather                  &      K. Arimatsu  \\
                                              & $35\degr$ $01'$ $31.0"$ N&    ZWO ASI-1600 Cool        &                            &       J. Takahash   \\
                                              & 449~m                               &                                   &                                   &           M. Tozuka    \\
                                              &                                      &                                   &                                   &            Y. Itoh         \\
                                             &                                      &                                   &                                   &                             \\
Okayama University           & $133\degr$ $55'$ $21.4"$ E &    $0.35~{\rm m}$       &     Bad weather               &   G. L. Hashimoto  \\
                                           & $34\degr$ $41'$ $19.1"$ N  &SBIG STL-1001E & &  M. Yamashita  \\
                                           &  42~m                                &                                   &                                   &                             \\
                                             &                              &                                   &                                   &                             \\
Bisei Spaceguard Center     & $133\degr$ $32'$ $40.0"$ E& $1.0~{\rm m}$         &      Bad weather              &   S. Urakawa   \\
                                              & $34\degr$ $40'$ $19.1"$ N&   mosaic CCD camera   &                                   &                             \\
                                              & 428~m    &                                   &                                   &                             \\
\enddata
\end{deluxetable*}

\clearpage
\begin{figure}[ht!]
\epsscale{1}
\plotone{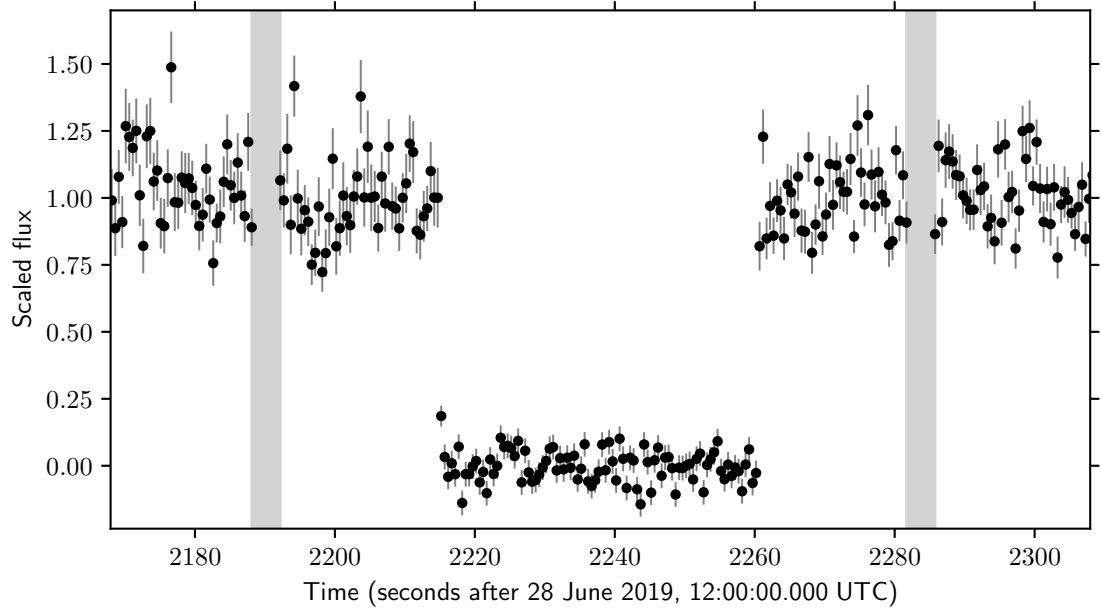}
\caption{Light curve of the occultation obtained with the Tomo-e Gozen camera.
The data are normalized to the unocculted flux.
An error bar for each data point represents the detector readout noise and target shot noise.
Gray regions are the time intervals between the individual data cubes' integrations (each corresponds to approximately 3.6~s).
}
\end{figure}

\clearpage
\begin{figure}[ht!]
\epsscale{1}
\plotone{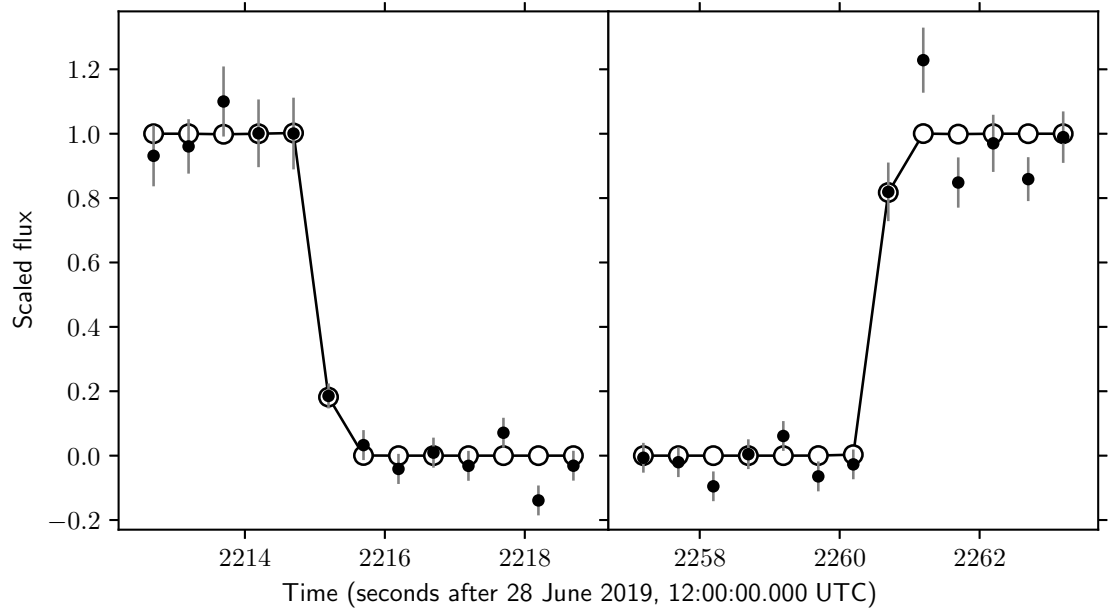}
\caption{The ingress (left panel) and egress (right panel) light curves overlaid with the best-fit synthetic light curve of a sharp edge occultation model.
The dots with error bars represent the individual observed data points, 
and the open circles represent the best fit model convolved by Fresnel diffraction, 
the angular diameter and the spectral energy distribution of the occulted star,
and the efficiency and the finite integration time of the Tomo-e CMOS sensor. 
See text for details.}
\end{figure}

\clearpage
\begin{figure}[ht!]
\epsscale{1}
\plotone{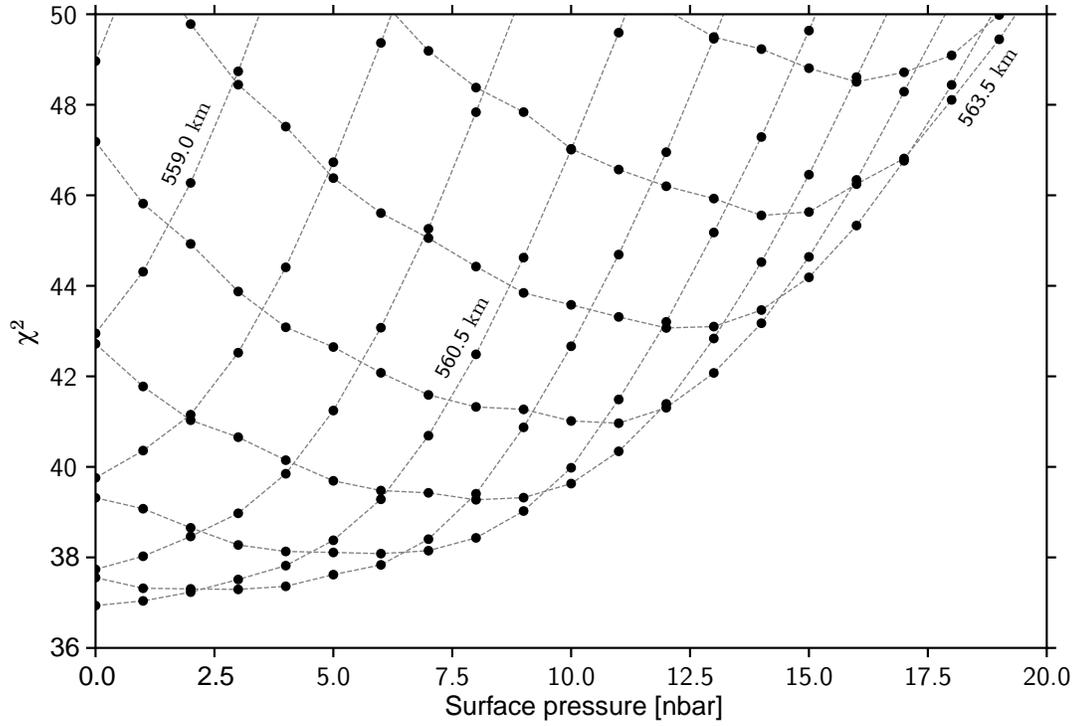}
\caption{Examples of $\chi^2$ values for different radii ($558.5 - 564.0 ~{\rm km}$) and surface pressures ($p_{\rm s} = 0 - 20 ~{\rm nbar}$).
In this model, each model assumes a pure ${\rm CH_4}$ atmosphere with a surface temperature, a temperature gradient, and an isothermal temperature above $\sim 10$ km are $T = 44~{\rm K}$, $\mathrm{d}T/\mathrm{d}z =5.7~{\rm K\, km^{-1}}$,  and $T = 102~{\rm K}$, respectively.
}
\end{figure}

\clearpage
\begin{figure}[ht!]
\epsscale{1}
\plotone{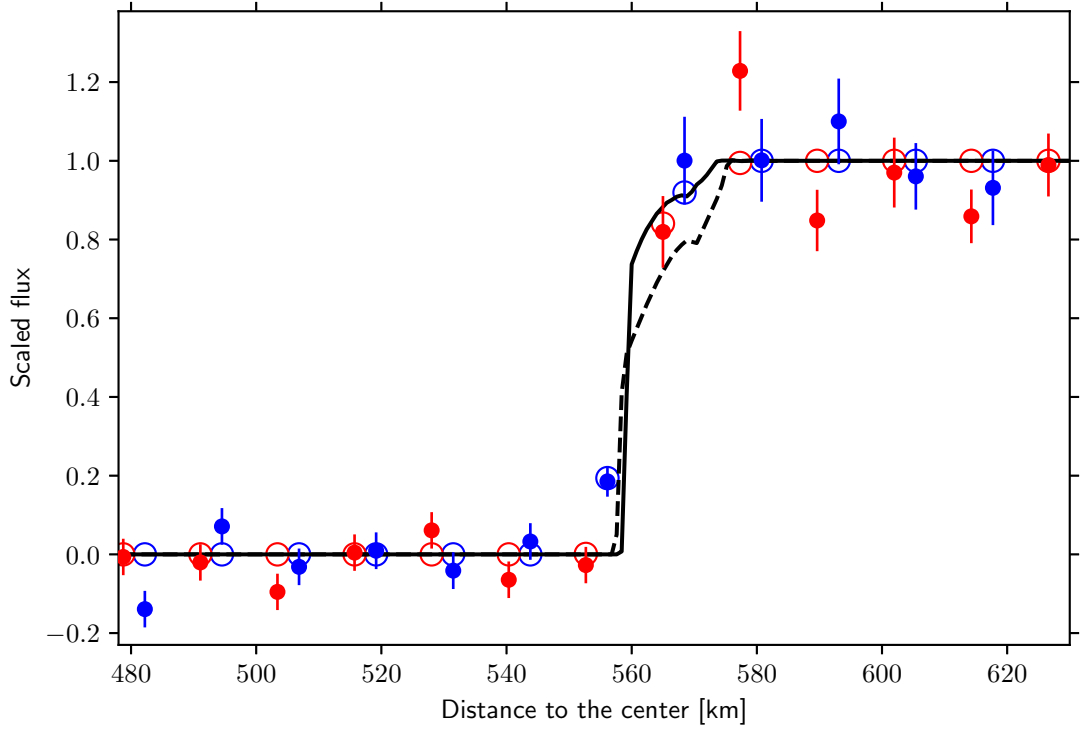}
\caption{The observed light curve (ingress and egress profiles shown as blue and red dots with errorbars, respectively) 
overlaid with the synthetic light curves of two ${\rm CH_4}$ pure atmospheric models.
The models correspond to a temperature profile increasing from $T = 44~{\rm K}$ at the surface to an isothermal $T = 102~{\rm K}$ above $\sim 10$ km with a thermal gradient $\mathrm{d}T/\mathrm{d}z =5.7~{\rm K\, km^{-1}}$. The solid and dashed lines represent the models with surface pressures of $p_{\rm s} = 6~{\rm nbar}$ and $p_{\rm s} = 16~{\rm nbar}$ corresponding to the $1\sigma$ and the $3 \sigma$ upper limits derived from the fit (see text for details), respectively.
Open circles correspond to the $p_{\rm s} = 6~{\rm nbar}$ synthetic light curve integrated over each bin.}
\end{figure}

\clearpage

\acknowledgments
We thank the anonymous referee for careful reading and providing constructive suggestions.
This research has been partly supported by Japan Society for the Promotion of Science (JSPS) Grants-in-Aid for Scientific Research (KAKENHI) Grant Numbers 15J10278, 16K17796, 18K13599, and 18K13606.
This research is also supported in part by the Japan Science and Technology (JST) Agencyfs Precursory Research for Embryonic Science and Technology (PRESTO), the Research Center for the Early Universe (RESCEU), of the School of Science at the University of Tokyo, and the Optical and Near-infrared Astronomy Inter-University Cooperation Program.

\bibliography{sample63}{}
\bibliographystyle{aasjournal}

\end{document}